\documentclass[aps, floatfix,
 reprint,
 amsmath,amssymb
]{revtex4-2}

\usepackage{graphicx}% Include figure files
\usepackage{dcolumn}% Align table columns on decimal point
\usepackage{bm}% bold math

\usepackage[colorlinks,linkcolor=blue,citecolor=blue]{hyperref}

\usepackage{url}  %Required

\begin{document}

%\preprint{APS/123-QED}

\title{Power Laws and Symmetries in a Minimal Model of Financial Market Economy}% Force line breaks with \\

\author{Liu Ziyin$^1$}
\author{Katsuya Ito$^2$}
\author{Kentaro Imajo$^2$}
\author{Kentaro Minami$^2$}
 %\altaffiliation{Physics Department, the University of Tokyo.}%Lines break automatically or can be forced with \\
\affiliation{$^1$Department of Physics, The University of Tokyo, 7-3-1 Hongo, Bunkyo-ku, Tokyo 113-0033, Japan\\
$^2$Preferred Networks, Inc., Otemachi Bldg. 1-6-1 Otemachi, Chiyoda-ku, Tokyo 100-0004, Japan 
}%


%\affiliation{
%\affiliation{
%\affiliation{%


\date{\today}% It is always \today, today,

\begin{abstract}
A financial market is a system resulting from the complex interaction between participants in a closed economy. We propose a minimal microscopic model of the financial market economy based on the real economy's symmetry constraint and minimality requirement. We solve the proposed model analytically in the mean-field regime, which shows that various kinds of universal power-law-like behaviors in the financial market may depend on one another, just like the critical exponents in physics. We then discuss the parameters in the proposed model, and we show that each parameter in our model can be related to measurable quantities in the real market, which enables us to discuss the cause of a few kinds of social and economic phenomena. %We also demonstrate the model's potential applicability by using it to simulate a stock market, exhibiting many realistic statistics. %. Additionally, we show that the model is flexible enough to be extended to many commonly studied theoretical scenarios in the field of econophysics and financial market modeling. 
%\begin{description}
%\end{description}
\end{abstract}

\maketitle

%\tableofcontents

\section{Introduction}
Modeling the real financial market has been a major challenge in the field of traditional economics \cite{sunder2006determinants, farmer2009economy} and in the emergent field of econophysics \cite{sornette2014physics, chakraborti2011econophysics, feigenbaum2003financial, toth2011anomalous, bucci2019crossover, kanazawa2018derivation, lux2001turbulence, lux1999scaling, levy1996power, kanazawa2018kinetic, laloux1999noise, mastromatteo2014anomalous}. The financial market can be modeled at a phenomenological level by writing down a stochastic differential equation for the price change, and this approach was first taken by Bachelier \cite{bachelier1900theorie}, a few years before Einstein's investigation of Brownian motion \cite{einstein1906theory}. The primary difficulty lies in modeling the interaction and behavior of individual participants of a financial market at the microscopic level \cite{sunder2006determinants, kanazawa2018derivation}. The classical economics approach assumes the rationality of human beings and that they maximize the predefined utility functions with some given information. The rationality assumption results in the Efficient Market Hypothesis, which predicts that the price of the stock market (or any market in general) follow a random walk and that, ultimately, the driving force of price change is exogenous, i.e., caused by an injection of new information \cite{fama1970efficient}. However, these predictions deviate far from what we observe in reality. For example, large price jumps occur about $7-8$ times per stock per day on average, but only $1$ news is released regarding each of the stocks every $3$ day \cite{joulin2008stock}; this suggests that external information is not sufficient to explain the market dynamics.

A series of universal statistical relations that could not be explained by classical economics are known to hold \cite{cont2001empirical} (known as the ``stylized facts"); one of the fundamental problems of socio-economic modeling is to explain the existence of these stylized facts \cite{levine2005finance}. For example, let $S_t$ denote the price of a stock at time $t$, then the return is defined as $r_t:=\log S_t /S_{t-1}$. The return is known to be heavy-tailed with kurtosis roughly equal to $4$, and, if one measures the c.d.f. of the largest returns of a stock, it is known to obey a power-law distribution with exponent roughly $-3$ to $-5$; the daily traded number of stock shares (called ``volumes") are known to have a power-law distribution with exponent $-3$ \cite{cont2001empirical}. What is more surprising about these facts is that these stylized facts appear almost universally across different nations, markets, and time \cite{goncu2018anatomy, panait2012stylized}; they even hold for the newly established Bitcoin market \cite{bariviera2017some}. This signature of universality calls for an explanation, while no consistent model exists yet to unify these phenomena. In fact, it might even be a question whether such a unifying model could exist. This situation is somewhat similar to the situation of turbulence in fluid dynamics. The first-principle theories following directly from the Navier-Stokes equation cannot explain the emergence of turbulence yet. However, it is widely expected that an ultimate correct theory needs to explain the well-known observed Kolmogorov $5/3$ power law \cite{kolmogorov1991local, barenblatt2003scaling} (along with a few other universal empirical facts).

In contrast to the classical economic models, the econophysics' approaches to the problem are often microscopically oriented and with phenomenology as high-level guidance. The line of work closest to physics is the Ising spin-based model by \cite{cont1997herd} and \cite{bornholdt2001expectation}. These models are based on the classical Ising model, and the main results are obtained by relabeling the physical objects to the relevant economic terms. For example, a spin is interpreted as a single investor, and the pairwise interaction between two spins is interpreted as the tendency of people to be influenced by other people around them. This line of work has been further developed to model even more complicated interactions between the investors \cite{paluch2015hierarchical}. However, the major criticism of these models is that they are unlikely to be realistic models of the markets and the agents constituting them. For example, the action of an agent is unlikely to behave like a spin, which only purchases $+1$ or $-1$ unit of stock at any given time. For detailed reviews and comparisons of the above-discussed models, see \cite{kukacka2020complex, barde2016direct, brock1998heterogeneous}. There are also microscopically founded models in the standard economics literature \cite{brock1998heterogeneous, gaunersdorfer2007nonlinear, gilli2003global, franke2012structural}, often called the ``heterogeneous agent-based models". However, these works have a vastly different purpose from the present work; they mostly focus on studying the agents' behavior in the specified model of the market and, thereby, explaining the market crashes and boom. The present work's goal is to explain the universality of many stylized facts. 

We argue that one major limitation of previous works is the lack of first-principle modeling of the cash flow. In a real economy, the stock price (or the price of any commercial goods) must be decided by the cash flow into and out of this stock. This cash flow out of the stock to the hands of the individual participants, and they must then reinvest based on their present wealth. A key component of first-principle modeling of the financial market must then be the modeling of cash flow, and the total wealth of the economy should be conserved if there is no explicit money-printing process. We, therefore, explicitly model the cash flow between the wealth of the individuals to the market, with partial conservation of money. Moreover, inspired by the scaling phenomena in physics, we argue that the existence of the universal relations in finance and economics may be understood through a similar mechanism in physics, where symmetries may lead to universalities. Therefore, we propose to establish a simple model of a financial economy, motivated by (1) symmetry constraints and (2) the requirement of minimality.

In summary, this work proposes a minimal model of a closed economy where the participants trade in a financial market, motivated by the real economic system's symmetries and microscopic necessities. From this minimal model, we see that the commonly observed universal scaling laws appear naturally and robust to initial conditions changes. The model's parameters can be related to measurable statistics of the real economy, increasing the explanatory and predictive power of the proposed theory. The organization of this paper is as follows. In the next section, we introduce the most directly related model to the proposed model, the limitation of which we use to motivate the present work. In section~\ref{sec: proposed model}, we establish the proposed model. In section~\ref{sec: theoretical properties}, we analyze our model in the mean-field regime. In Section~\ref{sec: discussion}, we link the model to the real economy and discuss the insights the model may provide to enhance our understanding of the economy. In Appendix F, we also demonstrate the usefulness of the proposed model by showing that it may be used as a computation model for simulating realistic financial data \footnote{See Supplemental Material at [URL will be inserted by
publisher] for the Appendix, which contains the numerical simulations and the derivations.}.

\begin{figure*}
    \centering
    \includegraphics[width=0.9\linewidth]{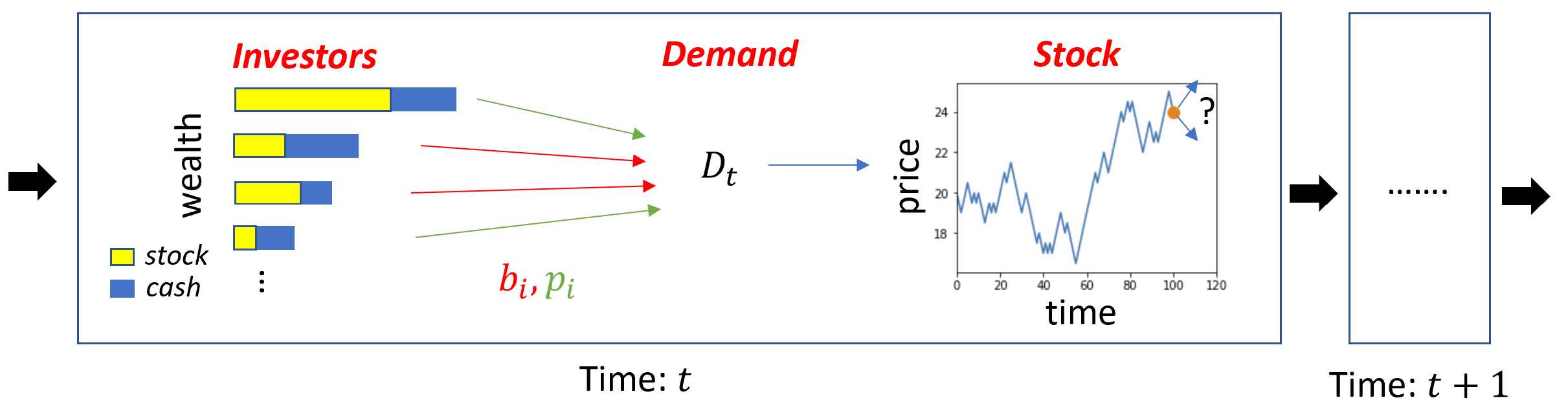}
    \caption{An illustration of the dynamics of a market. The market consists of the \textit{price} of a stock and a number of \textit{investors} (also called \textit{agents} and \textit{participants} in this work). At every time step $t$, the investors make financial decisions based on their wealth and the observed price. These decisions, taken together, create a net demand. The net demand creates a motion in the stock price. Previously, the dynamics of the wealth and the dynamics of the price have been separately modeled in a phenomenological way, and the main contribution of this work is to propose a minimal model that connects the two kinds of dynamics.}
    \label{fig:illustration market}
\end{figure*}
\section{Background and the Basic Models}\label{sec: background}
%[find work that is as close to the present work as possible]
\textit{Dynamics of price}. Finance is the study of the market. Two core concepts in finance is \textit{time} and \textit{uncertainty}. Time relates to the fact that the market is dynamic, i.e., changing with time. For example, the average price of a stock is different today from next month. The uncertainty aspect refers to the fact that the future price of anything cannot be determined for sure, i.e., knowing all the past prices and all the relevant information does not suffice to predict the price tomorrow with $100\%$ certainty. The combination of these two fundamental facts results in a (conjectured) consequence that the price of any product in the market obeys some stochastic differential equation (SDE):
\begin{equation}
    d S_t = f(S_t) dt  + dB_t
\end{equation}
for some random process $dB_t$. Alternatively, the price may obey a stochastic difference equation:
\begin{equation}
    \Delta S_t = f(S_t) \Delta t  + \eta_t
\end{equation}
for some random noise $\eta_t$. One picture assumes that the fundamental process of price change is continuous in time; the other assumes discreteness. So far, there is no clear favoring of one over another in the literature since it is unclear whether the fundamental price formation process is continuous-time or discrete-time (or neither). This indeterminacy is a consequence of the following observation: while the transactions may take place at any point in time, the transactions need to occur in a discrete manner. One cannot make a ``continuous" money transaction. 

Two most commonly adopted equations to describe the motion of a stock price is (1) the geometric Brownian motion (GBM)
\begin{equation}\label{eq: geometric brownian motion}
    S_{t+1} = (1 +r_s) S_t + \sigma_t S_t \eta_t
\end{equation} 
where $\eta_t$ is drawn from a Gaussian distribution; here, the word ``geometric" refers to the fact that the noise term $\sigma_t S_t \eta_t$ is proportional to the price $S_t$ itself; the other model is more recent, often called the Heston model \cite{heston1993closed}:
\begin{equation}\label{eq: heston model 1}
    \begin{cases}
        dS_t  = \mu S_t dt  + \sqrt{\nu_t} S_t dW_t;\\
        d\nu_t  = \kappa(\theta - \nu_t) dt  + \xi \sqrt{\nu_t} dU_t
    \end{cases}
\end{equation}
where $\nu_t$ is the instantaneous volatility. Unlike the geometric Brownian motion model, the Heston model assumes that the variance $\nu_t$ of the noise is also random, and follows a random walk of its own. These two models are the widely adopted phenomenological models of the price motion. The limitation of these two equations are obvious: (1) the investors are coarse-grained into the noise term and do not appear in the equation explicitly, and (2) the deterministic part of the equation is a simple linear dynamics, which cannot reflect the complicated non-equilibrium dynamics that the market is going through. 

\textit{Dynamics of Wealth}. There has also been strong interest in modeling the wealth dynamics of the individuals. There is a strong sense of how physics in general and thermodynamics in particular should be relevant to this problem because the wealth, like energy, should be conserved in total despite complicated microscopic exchanges of money between the investors \cite{Yakovenko_2009, dragulescu2000statistical}. Two basic phenomenological models of the evolution of wealth in an economy is proposed by \cite{bouchaud2000wealth, solomon2001power}. Ref.~\cite{solomon2001power} starts from the generalized Lotka--Volterra model while Ref.~\cite{bouchaud2000wealth} models the dynamics through an Ising spin model. Both approaches run into the problem that the full model is not analytically solvable, and the simplifying approach assumes that all the participants in the economy are identical and that they feel a static force; this approximation is, in essence, a mean-field approach, as is also used in \cite{bouchaud2000wealth}. In the mean-field limit, these two models reduce to a similar form:
\begin{equation}
    \frac{dW_i}{dt} = -a_0 W_i + a_1 + \sigma W_i \epsilon(t),
\end{equation}
where $W_i$ denotes the wealth of the $i$th participant in the economy. %The parameters $a_0$ and $a_1$ are given different interpretations. 
For \cite{bouchaud2000wealth}, $a_0$ is proportional to the rate of monetary exchange, $J$, between two different individuals, and $a_1$ is proportional to $J$ times the average wealth in the economy; equivalently, $a_0 W_i$ models the amount of money that the $i$th participant pays to other participants, while $a_1$ models the payment that $W_i$ receives from others. In this view, the connectivity of the underlying Ising model and the rate of exchange $J$ is crucial since a phase transition from an egalitarian society to a highly unequal society happens in the same way that a phase transition occurs in the standard Ising model when the disordered phase transitions to an ordered phase as the temperature are reduced. In \cite{solomon2001power}, the interpretation is similar; $a_1$ is interpreted as the regular income received by a person, and $a_0$ is the redistribution of wealth due to taxation, subsidies, and other fixed economic activities.

Given the Langevin equation above, the way to proceed is to write its equivalent Fokker-Planck equation and solve the stable distribution. Treating $a_0,\ a_1$ as constants \footnote{The calculation in the original work is actually about the relative wealth and is slightly more complicated in computation, but the ultimate interpretation does not change. To make these works more easily compared with the present work, we solve this slightly simplified version of the original.}, it is easy to show that the stationary distribution is
\begin{equation}
    P(W_i)\sim W_i^{-1 + \frac{a_0}{\sigma^2 }}\exp\left(-\frac{a_1}{\sigma^2 W_i}\right),
\end{equation}
and the exponent $a$ is the desired exponent of wealth. One limitation of these approaches is that the key parameter $a_0$ is not an observable. $a_0$ is proportional to the average rate of economic contact between participants in the market, but there is no way to measure this degree of contact objectively. When interpreted as wealth redistribution by government policies, it is far from clear how $a_0$ may be calculated unambiguously. The problem is exacerbated when one recognizes that the effect of fiscal policy on the economy is highly non-linear and that a first-order expansion in $W_i$ is insufficient to model such policies \cite{grossman1988government, brinca2019nonlinear}. We attribute the difficulty of making sense of the parameter $a_0$ to the lack of the specification of the underlying economic model in these models. One cannot model economic exchanges correctly without specifying how the transactions between investors occur. Therefore, we argue that one crucial step in the development of a first-principle theory of the market dynamics is the development of a model that consistently connects the dynamics of the price and the dynamics of the wealth. See Figure~\ref{fig:illustration market} for an illustration of the market dynamics we propose to understand.

We propose to model the investors' wealth and the dynamics of the market in a consistent framework. Some other attempts in this direction also exist \cite{levy1994microscopic, chiarella2006asset}, where the price and the wealth dynamics are modeled simultaneously; however, these models often have a large number of parameters, and the goal is to understand the population dynamics of the investors in market crashes and booms. In contrast, this work aims at establishing a minimal model to identify symmetries' role in forming generic and universal patterns in a financial market economy.

%\begin{enumerate}
%    \item motivate why classical economics does not work
%    \item motivate why a minimal model is needed
%    \item motivate why we need to link agent statistics and market statistics
%\end{enumerate}

%\clearpage

%{\small
\begin{table*}[t!]

    \centering
    \begin{tabular}{c|c|c|c}
    \hline\hline
        $\lambda,\ \sigma$ & Symmetry & Price Distribution & Wealth distribution  ($W \gg 1$)\\
    \hline
        $ \lambda=\lambda_0,\ \sigma=\sigma_0$ &
        $-$&
        $S^{\frac{b(2 M_0 + \lambda S_0^2)}{\lambda \sigma^2}} e^{f_1(S)}$ %e^{-\frac{ \left({b}+2p \right) }{ 2\sigma^2}\left[  S - \frac{p(\lambda S_0 - \Pi_0)}{b+2p}\right]^2}$
        &  $W^{\frac{b(2 M_0 + \lambda S_0^2)}{2\lambda \sigma^2} - \frac{1}{2}}$\\
        
       % $ \lambda=\lambda_0,\ \sigma=\sigma_0 S$ & 
    %    $-$&
     %   $S^{ -1 -\frac{2}{\sigma^2} \left(\frac{b}{2}+p \right)} e^{ \frac{2}{\sigma^2}\left[ -\frac{b(2 M_0 + \lambda S_0^2)}{4\lambda} S^{-2} + p\left(\frac{\Pi_0}{\lambda} -  S_0\right) S^{-1} \right]} $  \\

        $ \lambda=\frac{\lambda_0}{S},\ \sigma=\sigma_0 S $ &
        $S\to kS$& Not Power Law
        %$S^{-1 - \frac{2}{\sigma^2}\left[\frac{b}{2} + p(1 + \Pi_0) \right]}e^{f_2(S)}$ %e^{- \frac{2}{\sigma^2}\left[\frac{b(2 M_0 + \lambda S_0^2)}{2\lambda}S^{-1} + \left(\frac{b}{2}+p \right) S\right]} $ 
        
        & Not Power Law\\
        
        $ \lambda={\lambda_0 \Pi},\ \sigma=\sigma_0 $ & $\Pi \to k\Pi$& 
        Not Power Law
        %$S^{\frac{bS_0^2}{2}}e^{f_3(S)}$ %e^{2\frac{2}{\sigma^2} \left[\alpha S^{-\lambda} +\beta S^2 + \gamma S^{1-\lambda} - \kappa S^{1-\lambda} \right]}$
        & Not Power Law\\
        
        $ \lambda=\frac{\lambda_0 \Pi}{ S},\ \sigma=\sigma_0 S$ & 
        $S\to kS,\ \Pi\to k\Pi$
        &
        $S^{ -1 -\frac{2}{\sigma^2} \left(\frac{b}{\lambda_0 +1} + \frac{p}{\lambda_0}  \right)} e^{f_4(S)}$ %e^{ -\frac{2}{\sigma^2}\left[\alpha S^{-(1+\lambda_0)} +\beta S^{-2} + \gamma S^{-1} - \kappa S^{-\lambda_0}  \right]}$
        &  $W^{-1 - \frac{2}{\sigma^2 (\lambda_0 +1)}\left(\frac{b}{\lambda_0 +1} + \frac{p}{\lambda_0}  \right) }$\\

        \hline\hline
    \end{tabular}
    \caption{Distributions of price and wealth when \eqref{eq: stochastic main} obeys different kinds of symmetry. When both rescaling symmetries are satisfied, one obtains meaningful predictions for the price and wealth; universal power-law scaling that is initial-condition-independent only emerges when both symmetries are modeled for in the dynamical equation. Here, $b$ is the average tendency of buying, $p$ is the average tendency of selling, and $\lambda_0$ is the susceptibility of price to an excessive demand.}
    \label{tab:theory summary}
\end{table*}

\section{The Proposed Model}\label{sec: proposed model}
This section establishes our model, justified by empirical facts and fundamental symmetries in the problem. Let there be a single stock with price $S(t)$ and $N$ investors (agents) in the market. The complete state of the $i$th agent is given by a pair of scalars $M_i(t),\ \Pi_i(t) $, where $M_i$ is the amount of money he or she holds and $\Pi_i(t)$ the number of shares. Note that standard models such as the New-Keynesian models often assume a setting where a risky stock exists alongside a risk-less fixed-return government bond. However, this two-stock model is only meaningful when one incorporates a utility maximization procedure done in the standard economic models \cite{levy1994microscopic, FINOCCHIARO2011146, dieci2018heterogeneous}. Our framework can be extended to treat this setting, but this adds unjustified complication to the model, reducing the analytical tractability and interpretive power that we are attempting to achieve in the discussion. 

At any time $t\in \mathbb{Z}^+$, there are two possible actions for each agent: $b_i(t), \ p_i(t) \in  [0, 1]$ that denote the percentage of money the agent will invest in the stock and the percentage of share the agent will sell, respectively. Note that $b,\ p$ are unitless. A key feature that differentiates our model from the previous models is that we parametrize agents' actions by the unitless variables $b$ and $p$ that denote the willingness of the buying and selling, independent of how much money the agent has. This choice of $b$ and $p$ is natural to investment since it does not cause any income effect \footnote{For example, it does not seem reasonable to assume that richer people will invest a higher fraction of their wealth into the market in comparison to poorer people.}. After the transaction at time $t$, the change in the cash that $i-$th agent holds is the the return due to the stock sale minus the money spent: $\Delta M_i= p_i(t)\Pi_i(t) S(t) - b_i(t) M_i(t)$; likewise, one can write down the change in one's stock holding $\Delta \Pi_i$.
Therefore, at a given time $t$, the change in money and stock holding for the $i$th agent is then
\begin{equation}
    \begin{cases}
        \Delta M_i(t) = p_i(t)\Pi_i(t) S(t) - b_i(t) M_i(t) \\
        \Delta \Pi_i(t) = - p_i(t) \Pi_i(t) + \frac{b_i(t)M_i(t)}{S(t)}.
    \end{cases}
\end{equation}
Note that the above equations are correct by definition, and what is hard to model is the price change to close the system of equations above. Also, there has been strong evidence that setting $b_i, p_i$ to be random variables may be appropriate here.
It is discovered that when incorrect financial operations (such as spending more money than one has) are forbidden, computer agents making random financial decisions produce a price trajectory that is very close to a trajectory produced by real human beings \cite{gode1993allocative}. In the classical economics literature, such a choice is called zero-intelligence agents \cite{ladley2012zero, gode1993allocative}. Therefore, a model based on zero-intelligence traders offers a strong explanatory power because it allows one to identify whether an observed socio-economic phenomenon is caused by human rationality or by the fundamental properties of the market mechanisms.  

We assume that the change in price $\Delta S_t$ is an analytical function of all the observables in the market, such as the price and order histories. %all the past prices and orders. %(and also of external forces such as the arrival of new information).
The price change thus takes the following general functional form:
\begin{equation}
    \Delta S(t) = g\left[\{p(t') \Pi(t'), b(t') M(t'), S(t')  \}_{t'=1}^t \right],
\end{equation}
i.e., $S(t)$ can be a general functional. It follows from standard classical economic arguments the price is determined by the supply and demand curve of the market \cite{kyle1985continuous}. We therefore expand $S(t)$ to the first order in the immediate average demand $D(t):= \frac{1}{N} \sum_{i=1}^N \Delta \Pi_i(t)$:
%\begin{equation}
    $\Delta S(t) = c_0 - \frac{1}{\lambda} D(t) + O(D^2)$; %\end{equation},
for some positive constant $\lambda$, which can be called the market depth \cite{cont1997herd} and is the susceptibility of the price a perturbative external demand. Moreover, the price should not change when there is no net excessive demand in the market, and so $c_0=0$, and we arrive at our price change function:
\begin{equation}
    \Delta S(t) = - \frac{1}{\lambda} D(t); %+\sigma \epsilon(t);
\end{equation}
Note that $\lambda$ may still be a function of the observables in the market; for example, they may both depend on price.

While the price change function we use is linear, it generates non-trivial dynamics. Similar linear price impact functions can be derived from classical economic arguments \cite{kyle1985continuous}, and has been used in the Ising spin-based market model \cite{cont1997herd}. An increasing number of research studies the exact functional form of the price impact \cite{toth2011anomalous}; these interesting price impact functions can be studied in our framework by simply redefining the price impact function above. In this work, we use the linear price impact function. Therefore, the following set of equations determine our model of the market:
\begin{equation}\label{eq: main model}
    \begin{cases}
        \Delta M_i(t) = p_i(t)\Pi_i(t) S(t) - b_i(t) M_i(t);\\
        \Delta \Pi_i(t) = - p_i(t) \Pi_i(t) + \frac{b_i(t)M_i(t)}{S(t)};\\
        \Delta S(t) =  \frac{1}{\lambda} \sum_{i=1}^N \Delta \Pi_i(t).% + \sigma \epsilon(t).
    \end{cases}
\end{equation}
Note that, in this model, total wealth is only partially conserved. The first two equations conserve the total wealth, while the third line breaks such conservation \footnote{One popular way to enforce total wealth conservation is to define the price impact function through a Walrasian auction, but the Walrasian auction is, in itself, a highly non-linear mechanism and involves making unjustifiable about the price-demand curve of each agent.}. However, we do not think this is a significant problem because, in the real economy, it is never the case that the wealth and the financial market constitute a closed system, and part of the wealth may be distributed to other non-financial objects. In physics terms, one might imagine the existence of a conceptual heat-bath of money. While the total money of the system and the bath should be conserved, it is not the case that the wealth in the system is conserved. This also reflects the difficulty of determining the price impact function and that better modeling of $\Delta S$ will be crucial for future research.

Despite the formal simplicity of Eq.~\eqref{eq: main model}, $\lambda$ needs to depend on the price and other parameters. In this work, we assume that both the cash and the shares are infinitesimally divisible, which is a standard assumption in theoretical finance. Under this assumption, two fundamental symmetries exist in a financial market \footnote{On the other hand, we note that these two symmetries are broken if either the money or the share are not infinitesimally divisible, which is the case in reality. For example, in any monetary systems, the smallest amount of usable cash is lower bounded. Also, in real financial markets, the smallest amount of buyable shares are also lower-bounded. Our assumption amounts to assuming that these two bounding effects are negligible. This assumptions is also empirically justified. Taking the stock price of Apple as an example, the current price of Apple per stock is roughly $200\$$ per share, while the smallest unit of USD is $0.01\$$. Also, the daily average traded volume of Apple is roughly $10^8$ volume, orders of magnitudes larger than the minimum tradeable volume.}: (1) our decisions and the market should be unaffected by a rescaling of the unit of money $M\to kM$ and $S\to kS$ for some $k>0$ and, therefore, the dynamical equations should be invariant to such rescaling. This rescaling symmetry has been used previously in \cite{solomon2001power, bouchaud2000wealth} to derive the equation of motion of the wealth distribution but has not been applied to market modeling before; (2) likewise, the financial market should be invariant to a redefinition of the unit of the shares: $\Pi \to z\Pi$, $S \to S/z$ for some $z>0$. While the first two lines in Eq.~\eqref{eq: main model} are invariant to such rescaling, the third line is not for an arbitrary $\lambda$. Therefore, these two facts present additional functional constraints in the form of $\lambda$. One way to impose such a constraint is by defining 
\begin{equation}
    \frac{1}{\lambda} = \frac{S(t)}{\lambda_0 \Pi(t)}, %\quad \sigma= S(t)\sigma_0
\end{equation}
where $\Pi(t) := \frac{1}{N}\sum_i^N \Pi_i(t)$ is the average holding of the stock share at time $t$. Additionally, we are also also interested in the wealth, $W_i(t) := M_i(t) + S(t)\Pi_i(t)$, of the individuals and its stationary distribution $p(W_i)$.

%}

\section{Mean-Field Analysis}\label{sec: theoretical properties}

It is well-known that coupled sets of differential equations are difficult to analyze, and this is also a major challenge in the field of financial modeling; in some sense, the lack of analytical tractability is a major limitation of many of the financial models in the field \cite{chiarella2006asset}. The lack of an analytical solution limits the models' explanatory power. This problem also exists in the model we proposed in Eq.~\eqref{eq: main model}. We thus limit our theoretical study to the following minimalistic mean-field limit. We will see that, even in this simple mean-field analysis, the model already exhibits rich behavior.

The agents defined in the previous section take arbitrary strategies $b_i(t), p_i(t)$, which may result in arbitrarily complex interaction and price dynamics. The simplest choice for such strategies is a time-independent strategy. We also set $b_i, p_i$ to be random variables (i.e., we take the zero-intelligence limit). We proceed further by taking the mean field limit, where $b_i(t)=b$ and $p_i(t)=p$ for all $i$; we also require the agents start from the same initial condition such that $M_i(0)= M_0$ and $\Pi_i(0) = \Pi_0$. %The system of equations needs to remain meaningful when $N\to \infty$, and, thus, we let $\lambda = N \lambda$. 
Now we take the continuous-time limit, which is equivalent to assuming that the market depth $\lambda$ is sufficiently large so that the price cannot change too drastically in a short period of time. For arbitrary quantity $X$, $\Delta X \to \dot{X} dt$ and $p \to pdt$ and $b\to b dt$; the set of equations become
\begin{equation}\label{eq: set of equation theory}
    \begin{cases}
        \dot{ M}(t) = p\Pi(t) S(t) - b M(t) \\
        \dot{\Pi}(t) = - p \Pi(t) + \frac{bM(t)}{S(t)}\\
        \dot{S}(t) =  \frac{1}{\lambda} \dot{\Pi}(t) %+ \sigma \epsilon(t),
    \end{cases}
\end{equation}
where we have assumed the factor of $N$ into the definition of $\lambda$ and removed the dependence on subscript $i$, because the agents are identical in the mean-field limit.

In general, $\lambda=\lambda(S)$ can be a function of the price and the set of differential equations may be written as a single differential equation for $S$ \footnote{See Appendix A.}:
\begin{align}\label{eq: non simplied sde}
    \dot{S}(t) &= \frac{b \left[M_0 - \int_0^tdt  \lambda\frac{dS}{dt} S(t) \right]}{\lambda S(t)}  - \frac{p}{\lambda} \left( \Pi_0 +  \int_0^t dt \lambda\frac{dS}{dt} \right) + {\sigma} \epsilon(t),
\end{align}
where, to account for the random nature of the price, we add a noise term $\sigma \epsilon(t)$ to the right and $\epsilon(t)\sim \mathcal{N}(0,1)$. $\sigma$ is the volatility of the price and may also be a function of price. This stochastic term models some fundamental uncertainty in the determination of the price and may be derived from by treating $b$ and $p$ as random variables in the zero-intelligence limit \footnote{Or it may be a result of injection of external information into the market.}.

Theoretically, we focus on studying the theoretical properties of \eqref{eq: non simplied sde} with different choices of $\lambda$ and $\sigma$, determined through symmetry constraints. In this section, we also use the integrals of the deterministic equations as the definition of the other relevant variables; for example, $\Pi(t) := \Pi(0) +  \int_0^t dt \dot{S}/\lambda $.
The distribution for $S$ can be solved by first writing out the Fokker-Planck equation and finding the stationary distribution. We then take the set of equations in \eqref{eq: set of equation theory} as the definition for other quantities to find the distribution of wealth and return. We present the calculation in the appendix. We show the solutions for the most representative examples in Table~\ref{tab:theory summary}. We consider four different choices: (1) $\lambda=\lambda_0,\ \sigma=\sigma_0$; this is the case when no symmetry exists in the system. We see that it results in an unrealistic distribution for the price and wealth. The exponent for the power-law part has positive exponent, while the real world distributions always have negative exponents; (2) $\lambda=\lambda_0/S(t),\ \sigma=\sigma_0S(t)$: this system is invariant to a rescaling of the unit of money (denoted as the $S\to kS$ symmetry), and no explicit power-law behavior emerges in this case; (3) $\lambda=\lambda_0 \Pi(t),\ \sigma=\sigma_0$; this system is symmetric to redefining the unit of stock shares; again, one does not see the emergence of universal scaling behavior; (4) $\lambda=\lambda_0 \Pi(t)/ S(t),\ \sigma=\sigma_0S(t)$; this is the simplest kind of model that is invariant to both the redefinition of the unit of money and of share; we see that, surprisingly (or not), unitless scaling laws emerge for both the price and wealth distribution; more importantly, the exponents do not depend on the initial condition of the markets; this suggests the universality of these distributions. We discuss the meaning of the derived exponent in detail in the next section. Note that, when $\lambda = c_0 S^{c_1}$ (the fourth case), Eq.~\eqref{eq: stochastic main} simplifies to
\begin{multline}\label{eq: stochastic main}
   \dot{S}(t) = \frac{b\left[2 M_0 + \lambda' S_0^2/(c_1 +2)\right]}{2\lambda S(t)}  - \left(\frac{b}{c_1 + 2}+\frac{p}{c_1+1} \right) S(t) \\ - \frac{p}{\lambda} \left(\Pi_0 - \frac{\lambda' S_0}{c_1 + 1} \right) + {\sigma} \epsilon(t), 
\end{multline}
where we have defined a constant $\lambda':= c_0 S_0^{c_1}$.

\begin{table*}[t!]
    \centering
    \begin{tabular}{c|c|c|c}
    \hline\hline
        Observable & Pareto Exponent & Predicted Formula & Scaling Relation\\
        \hline
        Price $S(t)$  & $\alpha$ & $\alpha = \frac{2}{\sigma^2}\left(\frac{b}{\lambda_0 +1} + \frac{p}{\lambda_0}  \right)$ &  $\alpha $\\
        Wealth $W(t)$ & $\beta$  & $\beta = \frac{2}{\sigma_0^2(\lambda_0 +1)}\left(\frac{b}{\lambda_0 +1} + \frac{p}{\lambda_0}  \right)$ & $\frac{\alpha \gamma}{\alpha + \gamma}$\\
        Volume $V(t)$ & $\gamma$ & $\gamma = \frac{2}{\sigma^2\lambda_0}\left(\frac{b}{\lambda_0 +1} + \frac{p}{\lambda_0}  \right) $ & $\frac{\alpha}{\alpha -\beta}$\\
       % Return $R(t)$ & $\rho=$  & $\begin{cases}
       %     \frac{3}{2} - \frac{1}{\sigma_0^2}\left(\frac{b}{2} + p  \right) \\
        %    \frac{}{} 
        %\end{cases}$ & $\frac{1}{2}(3-\alpha)$\\
        \hline\hline
    \end{tabular}
    \caption{Pareto exponents of price, wealth, and traded stock volume and the relation between different exponents. We note that the model predicts the power-law exponents to be dependent on one another, which is reminiscent of the scaling relations in critical phenomena. Here, $b$ is the average tendency of buying, $p$ is the average tendency of selling, and $\lambda_0$ is the susceptibility of price to an excessive demand.}
    \label{tab:exponents}
\end{table*}

It is worth exploring the fourth model with both symmetries deeper. Besides the price and wealth distributions, stylized facts are also known to exist for the return $R_t :=\log(S_{t}/S_{t-1}) \approx \delta t \frac{d}{dt} \log(S) $, and the traded volume $V_t := \Delta \Pi_t \approx \delta t \dot{\Pi}(t)$. See Table~\ref{tab:exponents} for a summary of these quantities in our theory. We see that all price, wealth, and volume obey heavy tail distribution with unitless exponents. The mean-field theory not only predicts the formula for each of the quantities, but it also predicts two relations between them, which we depict in the fourth column. This relation is reminiscent of the scaling relations in critical phenomena. Moreover, our theory predicts that the Pareto exponent of wealth $\beta$ is always smaller than the exponent of volume and price.

\section{Implications}\label{sec: discussion}
While we explicitly referred to each agent as a person, one agent may also be interpreted as a collection of people who share the propensity for investment, e.g., an institution, a fund, or even the economy as a whole (when the mean-field limit is taken). Of particular interest here is when the agents are interpreted as the representative of the whole economy. In this interpretation, $b_i$ and $p_i$ become the economy's average tendency to buy and sell. This interpretation is especially appropriate for the mean-field model in Eq.~\eqref{eq: stochastic main} because all the agents are assumed to be identical. In this light, we discuss the implications of the proposed model.

We first link the model's parameters to real-world measurable quantities. Four unitless quantities, $\sigma_0,\ \lambda_0,\ b, \ p$, exist in our theory, and the scaling exponents predicted by our theory are dependent on them. The market depth $\lambda_0$ may be measured by measuring $\beta$ and $\gamma$:
\begin{equation}
    \lambda_0 = \frac{\gamma}{\gamma-\beta}.
\end{equation}
Since the market depth must be positive, we have that $\gamma \geq \beta$; this means that the tail of trading volume cannot be heavier than the tail of the wealth. This condition agrees with the intuition that the investment one makes cannot be larger than the amount of available wealth. For an economy, it is also true. At a macroscopic level, one expects $\gamma$ to be very close to its lower limit, i.e., $\gamma \approx\beta$ due to borrowing and leveraging money from the bank for investment. This agrees with the measured value in real economies where $\beta\approx 1.36$ and $\gamma \approx 1.40$ \cite{levy1997new, mantegna1995scaling}. Plugging in, we can estimate the value $\lambda_0\approx 35 \sim 10^1$, or $1/\lambda_0 = 0.029$. 

The parameters $b$ and $p$ can also be related to measurable quantities. %In fact, they are functions of 
We define the total market capitalization $\mu:=N\Pi S$ and the total wealth of the society $W_{tot} :=N(M + \Pi S)$. Close to a stationary state, the injection of money into the economy must be equal to that leaving the market; we thus have
\begin{equation}
    bNM = {pN\Pi S} \longrightarrow b = \frac{\mu}{W_{tot} - \mu} p :=\kappa p,
\end{equation}
which relates $b$ to $p$ through the measurable quantities $\mu$ and $W_{tot}$, and the quantity $\kappa$ may be called the ``market activity index": the higher the $\kappa$, the more active a market is. Data shows that \footnote{Data from Credit Suisse and Worldbank: https://www.credit-suisse.com/about-us/en/reports-research/global-wealth-report.html, {https://data.worldbank.org/indicator/CM.MKT.LCAP.CD}.}
, across different countries, $\kappa$ is a peaked distribution, centering around $0.25$. It is now useful to consider the meaning of the parameter $p$. $p$ is defined as the amount of stock sold through a unit time $\Delta t$; therefore, $p$ is directly linked to the stock's liquidity. When $\Delta t$ is taken to be a day, $p$ is the stock's daily turnover rate (or the daily equivalent of the annual average turnover rate), a well-measured quantity. We denote the daily turnover rate as $p^*$ from now on. The annual turnover rate for most countries centers around $1.30$, which translates to a daily rate of $0.0036$.
This means that one may rewrite the factor as $b/(\lambda_0 + 1)+p/\lambda_0= [\kappa/(\lambda_0 + 1)  + 1/\lambda_0]p^*$, which $\approx 0.0080$ on average in the world. One may obtain a different estimate of the market depth value from this analysis to check the consistency of the theory. Here, \begin{align*}
    \beta &= \frac{2}{\sigma^2 (\lambda_0 +1)}\left(\frac{\kappa}{\lambda_0 + 1}  + \frac{1}{\lambda_0} \right)p^* \\
    \longrightarrow \lambda_0 &\approx \sqrt{\frac{2(1 + \kappa)p^*}{\beta \sigma^2}} \approx 12 \sim 10^{1}, %\nonumber
\end{align*}
where we have approximated $\sigma_0$ by the daily stock price volatility, which is of order $0.01$ for the US stocks; this estimation of $\lambda_0$ agrees in the order of magnitude with the independent estimate of the market depth in the previous paragraph. Therefore, every parameter of our proposed model can be measured in real market and economies in principle, and more importantly, the theory gives consistent predictions regarding the market depth and the empirical power law exponents.

Now it is interesting to compare with the result on the distribution of wealth in \cite{bouchaud2000wealth} and \cite{solomon2001power}. In \cite{bouchaud2000wealth}, the Pareto exponent takes the form $a_0/\sigma^2$, where $a_0$ is proportional to the rate of monetary exchange in the economy and $\sigma^2$ is the degree of randomness in the wealth acquisition process. In comparison, this work predicts an exponent of $\frac{b+2p}{\sigma_0^2(\lambda_0 + 1)}$, and one can naturally see that the $\sigma^2$ terms are analogous, and the term $\frac{b+2p}{\lambda_0 + 1}$ translates directly to the term $a_0$ in the Cont--Bouchaud model; as discussed, the term $\frac{b+2p}{\lambda_0+1}$ is proportional to the market activity weighted by the sensitivity of the market to external stimulus, which may be called `rate of monetary exchange'. In this sense, this work gives the parameters in the Cont--Bouchaud model a precise meaning. Ref.~\cite{solomon2001power} avoids interpreting the $a_0/\sigma^2$ term directly, but links this term to the average number of members in a household $L$ and argue that the Pareto exponent is equal to $\alpha =\frac{L}{L-1}$; the Pareto index is a monotonically decreasing function of $L$; therefore, the larger the number of average members in a family, the more inequality exists in society. In our theory, we showed that $\alpha\sim \kappa p$ which is a measure of market activity; one interpretation is that the market activity is inversely proportional to the size of a household, which corroborates with the long observed fact that child birth rate is inversely correlated with the average income in society \cite{simon1969effect}.

One might also discuss the effect of economic growth or inflation on society within the framework. For example, we can model the effect of economic growth or inflation by adding a term, $f(S)$ to the right-hand side of Eq.~\eqref{eq: stochastic main}. We consider and compare two kinds of growth; one is constant $f(S)=k_0$: this only shifts the equilibrium value of the price and does not affect the distribution's power-law exponents. Alternatively, we might consider a growth proportional to price $f(S) = k_0 S$: this directly affects the exponents, and changes the term $\frac{b}{\lambda_0 + 1} + \frac{p}{\lambda_0}$ to $\frac{b}{\lambda_0 + 1} + \frac{p}{\lambda_0} - k_0$. Note that, in economic growth (or inflation), $k_0>0$, and, thus, growth increases inequality by lowering the Pareto exponent; on the other hand, economic decay reduces inequality. This implication agrees with the intuition that a growing market tends to create extremely rich people either by chance or through their better investment skills, and rich people may become richer through some self-reinforcing mechanism, for example, by becoming more influential to society.

%[economic acitivity and inequality]
%[entropy maximization]
%[implication for policy]
%[experiment]
%[implication: volatility clustering, more on stylized facts]

%[TODO, compare with Bouchaud 2000, and Solomon 2001]
%[maybe plot the volatility itself]

%[normalize autocorrelation]

%\small{

\section{Concluding Remark}\label{sec: conclusion}
In this work, we have argued from the basic principles what the simplest form of a financial market model should take: it needs to connect the price dynamics to the wealth dynamics. The dynamics we derived is based on fundamental symmetry constraints; when analyzed in a mean-field regime, the dynamics leads to interesting universal behaviors that mimic the real market. One insight that this work brings is that the universality of the empirically observed power-laws in the real market might emerge due to the relevant symmetries in the market. What is more, as in the standard physical critical phenomena, the scaling exponents are dependent on one another; this corroborates the success of dimension analysis for understanding economic and financial systems \cite{daniels2003quantitative}. 

However, the simple model and our analysis of it are minimal. For example, the analysis gives no sensible prediction about the statistics related to the return, nor does it answer what heterogeneous agents and their interaction may bring to the financial system. Nevertheless, this work may pave the way for further understanding the cause and the nature of the commonly observed stylized facts in economics, and it may also serve as a ``baseline" model for modeling financial processes in the industry. Possible future works include finding more connections between more power-law indices and, possibly, predicting the existence of yet unnoticed power laws. One might also investigate the influence of various fiscal-economic policies on the financial system. The price dynamics may also be investigated under the assumption of the existence of a fundamental price. As has been briefly remarked in this work, the thermodynamics of the proposed microscopic model should also be interesting to investigate; for example, a recent work showed how a fundamental relation in theoretical finance could be directly linked to stochastic thermodynamics \cite{ziyin2022universal}. It would be an exciting and important future work to identify more links between statistical physics and finance.

\begin{acknowledgments}
We need to thank many people for useful discussions during the writing of this work. Shota Imaki has provided many valuable comments regarding finance and economics. Takuya Shimada provided many useful discussions and advices in the process. Financially, Liu Ziyin is partially supported by the GSS scholarship of the University of Tokyo.
%%We wish to acknowledge the support of the author community in using
%REV\TeX{}, offering suggestions and encouragement, testing new versions,
\end{acknowledgments}

%\bibliography{ref}% 

\end{document}

\title{Appendix for ``\textit{Power Laws and Symmetries in a Simple Model of Financial Market Economy}"}
%\date{\vspace{-5ex}}

\author{Liu Ziyin$^1$, Katsuya Ito$^2$, Kentaro Imajo$^2$, Kentaro Minami$^2$\\
{\small \textit{$^1$Department of Physics, The University of Tokyo, 7-3-1 Hongo, Bunkyo-ku, Tokyo 113-0033, Japan}}\\
{\small\textit{$^2$Preferred Networks, Inc., Otemachi Bldg. 1-6-1 Otemachi, Chiyoda-ku, Tokyo 100-0004, Japan}} }
 %\altaffiliation{Physics Department, the University of Tokyo.}%Lines break automatically or can be forced with \\
 
\maketitle

\onecolumngrid

\appendix

\section{Derivation of the Langevin Equation}\label{app sec: langevin derivation}
\subsection{Constant $\lambda$}
The set of differential equations can be written as
\begin{equation}
    \begin{cases}
        \dot{S}(t) = \frac{1}{\lambda} \dot{\Pi}(t)\\
        \dot{M} = -S \dot{\Pi}(t) \\
        \dot{\Pi}(t) = \frac{bM(t)}{S(t)} - p\Pi(t).\\
    \end{cases}
\end{equation}
Plugging in the first and the second equation to the third, this leads to a differential equation for $S(t)$:
\begin{align}
    \dot{S}(t) &= \frac{b \left[M_0 - \int_0^tdt  \lambda\frac{dS}{dt} S(t) \right]}{\lambda S(t)}  - \frac{p}{\lambda} \left( \Pi_0 +  \int_0^t dt \lambda\frac{dS}{dt} \right).
\end{align}
When $\lambda$ is a constant in $S$, the above equation simplifies to
\begin{align}
    \dot{S}(t) &=\frac{1}{\lambda} \left[ \frac{b M_0}{S(t)} - \frac{b\lambda (S^2(t) - S_0^2)}{2S(t)} - p\Pi_0 - p \lambda(S(t) - S_0) \right]\\
    & = \frac{b(2 M_0 + \lambda S_0^2)}{2\lambda S(t)} - \left(\frac{b}{2}+p \right) S(t)  - \frac{p}{\lambda}(\Pi_0 - \lambda S_0) \label{eq: constant lambda diffeq}.
\end{align}

\subsection{$S$-dependent $\lambda$}
On the other hand, when $\lambda = c_0 S^{c_1}$ and $c_1\neq -1$, we have 
\begin{equation}
    \begin{cases}
        \int_0^tdt  \lambda\frac{dS}{dt} S(t) = \frac{c_0}{c_1 + 2}(S^{c_1 +2} - S_0^{c_1 + 2});\\
        \int_0^tdt  \lambda\frac{dS}{dt}  = \frac{c_0}{c_1 + 1}(S^{c_1 +1} - S_0^{c_1 + 1}).
    \end{cases}
\end{equation}
Therefore, 
\begin{align}
    \dot{S}(t) &= \frac{b \left[M_0 - \int_0^tdt  \lambda\frac{dS}{dt} S(t) \right]}{\lambda S(t)}  - \frac{p}{\lambda} \left( \Pi_0 +  \int_0^t dt \lambda\frac{dS}{dt} \right)\\
    &= \frac{1}{\lambda} \left[ \frac{b M_0}{S(t)} - \frac{b (\lambda S^2(t) - \lambda' S_0^2)}{2 (c_1+2) S(t)} - p\Pi_0 - \frac{p ( \lambda S(t) - \lambda' S_0)}{c_1 +1} \right] \\
    &=\frac{b\left[2 M_0 + \lambda' S_0^2/(c_1 +2)\right]}{2\lambda S(t)} - \left(\frac{b}{c_1 + 2}+\frac{p}{c_1+1} \right) S(t)  - \frac{p}{\lambda} \left(\Pi_0 - \frac{\lambda ' S_0}{c_1 + 1} \right),\label{eq: diffeq c1!=-1}
\end{align}
where $\lambda' := c_0 S_0^{c_1}$ is an initial condition-dependent constant. To this equation, we introduce the noise term $\sigma \epsilon(t)$:
\begin{align}
    \dot{S}(t) =\frac{b\left[2 M_0 + \lambda' S_0^2/(c_1 +2)\right]}{2\lambda S(t)} - \left(\frac{b}{c_1 + 2}+\frac{p}{c_1+1} \right) S(t)  - \frac{p}{\lambda} \left(\Pi_0 - \frac{\lambda' S_0}{c_1 + 1} \right) + \sigma \epsilon(t).
\end{align}
As we have mentioned, the noise term can be directly derived by defining $b$ and $p$ as time-dependent random variables in the zero-intelligence limit, but, in this work, we treat $\epsilon$ as a external term for generality and derive its form from symmetry constraints.

Lastly, when $c_1=-1$, we have 
\begin{equation}
    \begin{cases}
        \int_0^tdt  \lambda\frac{dS}{dt} S(t) = \frac{c_0}{c_1 + 2}(S^{c_1 +2} - S_0^{c_1 + 2}) = {c_0}(S - S_0) ;\\
        \int_0^tdt  \lambda\frac{dS}{dt}  = {c_0}\log (S/S_0).
    \end{cases}
\end{equation}

One thus obtains a different differential equation:
\begin{align}
    \dot{S}(t) &= \frac{b \left[M_0 - \int_0^tdt  \lambda\frac{dS}{dt} S(t) \right]}{\lambda S(t)}  - \frac{p}{\lambda} \left( \Pi_0 +  \int_0^t dt \lambda\frac{dS}{dt} \right)\\
    &= \frac{b [M_0 - c_0(S-S_0)]}{c_0}  - \frac{p}{\lambda} \left( \Pi_0 +  c_0 \log S/S_0 \right)\\
    &= b\left( \frac{M_0}{c_0} + S_0 \right) - \frac{p\Pi_0}{c_0 S} - \frac{p}{S} \log S/S_0 - bS. \label{eq: c1=-1}
 \end{align}

\section{Stationary distribution derivation}\label{app sec; fourth model derivation}

\subsection{Case: $\lambda = \frac{\lambda_0\Pi}{S}$ and $\sigma=\sigma_0 S$}
We focus on the case when $\lambda = \frac{\lambda_0\Pi}{S}$ and $\sigma=\sigma_0 S$, since this is the most interesting case. We follow the procedure in the main text, where we first write out the equivalent deterministic equation of motion for $S$ in this dynamics, and then add the noise term. Recall that the deterministic set of equations are 
\begin{equation}
    \begin{cases}
        \dot{ M}(t) = p\Pi(t) S(t) - b M(t) \\
        \dot{\Pi}(t) = - p \Pi(t) + \frac{bM(t)}{S(t)}\\
        \dot{S}(t) =  \frac{S(t)}{\lambda_0 \Pi(t)} \dot{\Pi}(t); %+ \sigma \epsilon(t),
    \end{cases}
\end{equation}
we first would like to write $\lambda$ as purely a function of $S(t)$. Solving the third equation, we obtain
\begin{equation}
    \Pi=\Pi_0 \left(\frac{S}{S_0}\right)^\lambda_0.
\end{equation}
This means that $\lambda = \frac{\lambda_0 \Pi_0  }{S_0^{\lambda_0}}S^{\lambda_0 -1}$, and $\lambda' = \lambda_0 \Pi_0/S_0$. This can then be plugged into Eq.~\eqref{eq: diffeq c1!=-1} to obtain the following Langevin equation we would like to solve:
\begin{align}
    \dot{S}(t) &= \frac{b\left[2 M_0 + \lambda' S_0^2/(c_1 +2)\right]}{2\lambda S(t)} - \left(\frac{b}{c_1 + 2}+\frac{p}{c_1+1} \right) S(t)  - \frac{p}{\lambda} \left(\Pi_0 - \frac{\lambda' S_0}{c_1 + 1} \right) +  {\sigma_0} S(t) \epsilon(t)  \nonumber\\
    & = \frac{b S_0^{\lambda_0}}{2\lambda_0 \Pi_0}\left( 2M_0 + \frac{\lambda_0 \Pi_0 S_0}{c_1 + 2}\right)S^{-\lambda_0}(t)  - \left(\frac{b}{\lambda_0 + 1}+\frac{p}{\lambda_0} \right)S(t) - \frac{pS_0^{\lambda_0}}{\lambda_0}\left(1 - \frac{\lambda_0}{c_1 + 1} \right) S^{1-\lambda_0}+  {\sigma_0} S(t) \epsilon(t) \nonumber\\
    & = \frac{b S_0^{\lambda_0}}{2\lambda_0 \Pi_0}\left( 2M_0 + \frac{\lambda_0 \Pi_0 S_0}{\lambda_0 + 1}\right)S^{-\lambda_0}(t)  - \left(\frac{b}{\lambda_0 + 1}+\frac{p}{\lambda_0} \right)S(t) +  {\sigma_0} S(t) \epsilon(t) ,
\end{align}
where the terms are arranged with increasing orders in $S(t)$ (assuming that $\lambda \geq 2$; as discussed in the text, the measured values of $\lambda$ is of order $10^1$). The equivalent potential is then 
\begin{equation}\label{eq: effective potential}
    U(S) =  \frac{b S_0^{\lambda_0}}{2\lambda_0(\lambda_0 -1 ) \Pi_0}\left( 2M_0 + \frac{\lambda_0 \Pi_0 S_0}{\lambda_0 + 1}\right)S^{-(\lambda_0 -1 )}(t) + \frac{1}{2}\left(\frac{b}{\lambda_0 +1}+\frac{p}{\lambda_0} \right) S^2.
\end{equation}
The equivalent Fokker-Planck equation \cite{bouchaud2000wealth, levy1996power} (or by transforming to $y:= \log S$) is 
\begin{equation}
    \frac{d}{dt}p(S,t) = \frac{d}{dS} \left[ p(S, t)\frac{d}{dS} U(S) \right] + \frac{\sigma_0^2}{2}\frac{d}{dS} S \frac{d}{dS}[SP(S,t)],
\end{equation}
whose stationary solution is
\[  \log P (S) = -\frac{2}{\sigma^2} \int dS \frac{1}{S^2} \left(\frac{d}{dS}U + \frac{\sigma^2}{2}S  \right).\]
The solution takes the form $e^{f(S)}S^{-c_2} $ for some constant $c_2$, and we care about the the exact value of $c_2$. When $U$ is a polynomial in $S$, we can let $c_3$ be the coefficient of the second order term of $U(S)$ and find $c_2$ to be 
\begin{equation}
    c_2 = 1 + \frac{4}{\sigma^2}c_3,
\end{equation}
whereas the coefficients of other terms enter the exponential term. From Eq.~\eqref{eq: effective potential}, we have that $c_2 = 1 + \frac{2}{\sigma^2}\left(\frac{b}{\lambda_0 +1}+\frac{p}{\lambda_0} \right)$.

\subsection{Distribution of wealth and volume}\label{app sec: distribution of wealth}
Now we are ready to calculate the distribution of wealth, $W=M+\Pi S$. Again, we first try to write $W$ as purely a function of $S(t)$, which leads to 
\begin{equation}
    W(S) = M_0 + 2 \frac{\Pi_0}{S_0^{\lambda}}S^{\lambda_0 + 1} \sim S^{\lambda_0 + 1}.
\end{equation}
We can now perform a transformation of variable to the distribution $p(S)$:
\begin{align}
    p(W)|_{W\gg 1} dW &\sim S^{-1 - \frac{2}{\sigma^2}\left(\frac{b}{\lambda_0 +1} + \frac{p}{\lambda_0}  \right)} dS \\ 
    &\sim W^{-1-\frac{2}{\sigma^2(\lambda_0 + 1)}\left(\frac{b}{\lambda_0 +1} + \frac{p}{\lambda_0} \right)} dW.
\end{align}

We now derive the distribution for the traded volume $V := \Delta \Pi \sim \delta t \dot{\Pi}$. It has been shown that, for daily level data, the distribution of the volume is independent of the time scale $\delta t$, and so we can safely treat $\delta t$ as a constant factor and ignore it \cite{mantegna1995scaling}. By definition, we have 
\begin{equation}
     \dot{S}(t) = \frac{1}{\lambda} \dot{\Pi}(t)
\end{equation}
and we have shown that $\lambda = \frac{\lambda_0 \Pi_0  }{S_0^{\lambda_0}}S^{\lambda_0 -1}$, and so
\begin{equation}
    \dot{\Pi}(t) \sim S^{\lambda_0 - 1} \dot{S}(t)|_{S\gg 1} \sim S^{\lambda_0},
\end{equation}
and we can apply the transformation rule to the above factor to obtain that 
\begin{align}
    P(V) dV &\sim  S^{-1 - \frac{2}{\sigma^2}\left(\frac{b}{\lambda_0 +1} + \frac{p}{\lambda_0}  \right)} dS  \nonumber\\
    &\sim V^{-1 - \frac{2}{\sigma^2 \lambda_0}\left(\frac{b}{\lambda_0 +1} + \frac{p}{\lambda_0}  \right)} dV.
\end{align}

\section{Case: $\lambda = {\lambda_0}$ and $\sigma=\sigma_0$}

In this section, we find the stationary distribution of the case when both $\lambda$ and $\sigma$ are constant. The SDE under consideration follows Eq.~\eqref{eq: constant lambda diffeq}:
\begin{equation}
    \dot{S} = \frac{b(2 M_0 + \lambda_0 S_0^2)}{2\lambda_0 S(t)} - \left(\frac{b}{2}+p \right) S(t)  - \frac{p}{\lambda}(\Pi_0 - \lambda_0 S_0) + \sigma_0 \epsilon(t).
\end{equation}
The equivalent Fokker-Planck equation (at stationarity) is:
\begin{equation}
    0 = \frac{d}{dS}\left[P(S)\frac{d}{dS} U(S)\right] + \frac{\sigma^2_0}{2} \frac{d^2}{dS^2} P(S),
\end{equation}
whose solution is precisely the Boltzmann distribution (because the noise is thermal):
\begin{equation}
    P(S) \sim e^{-\frac{2U(S)}{\sigma_0^2}}.
\end{equation}
The power-law term in $P(S)$ comes from the logarithmic term in $U(S)$, whose coefficient is $\frac{b(2 M_0 + \lambda_0 S_0^2)}{2\lambda_0}$, and thus the power-law exponent for the stationary distribution is
\begin{equation}
    \frac{b(2 M_0 + \lambda_0 S_0^2)}{2\lambda_0}.
\end{equation}
Following the same calculation as in Section~\ref{app sec: distribution of wealth}, one can also find the power-law exponent of wealth as given in Table 1 of the main text.

\section{Case: $\lambda = {\lambda_0 / S}$ and $\sigma=\sigma_0 S$}

Here, $\lambda = \lambda_0/S$, and $\lambda' = \lambda_0 /S_0$. This can then be plugged into Eq.~\eqref{eq: c1=-1} to obtain the generalized Langevin equation we would like to solve:
\begin{align}
    \dot{S}(t) &=  b\left( \frac{M_0}{\lambda_0} + S_0 \right) - \frac{p\Pi_0}{\lambda_0 S} - \frac{p}{S} \log S/S_0 - bS +  {\sigma_0} S(t) \epsilon(t)\\
    &= - \frac{d}{dS} U(S) + {\sigma_0} S(t) \epsilon(t)
\end{align}
%The equivalent potential is then 
%\begin{equation}\label{eq: effective potential c1=-1}
%    U(S) =   b\left( \frac{M_0}{\lambda_0} + S_0 \right) - \frac{p\Pi_0}{\lambda_0} - \frac{p}{S} \log S/S_0 - \frac{1}{2}bS^2
%\end{equation}
The equivalent Fokker-Planck equation is 
\begin{equation}
    \frac{d}{dt}p(S,t) = \frac{d}{dS} \left[ p(S, t)\frac{d}{dS} U(S) \right] + \frac{\sigma^2}{2}\frac{d}{dS} S \frac{d}{dS}[SP(S,t)],
\end{equation}
whose stationary solution is
\[  \log P (S) = -\frac{2}{\sigma^2} \int dS \frac{1}{S^2} \left(\frac{d}{dS}U + \frac{\sigma^2}{2}S  \right), \]
The solution takes the form $e^{f(S)}S^{-c_2 + g(x)} $ for some constant $c_2$ and some polynomial $g$. The $g(x)$ polynomial is due to the existence of the $\log$ term in the effective potential and this means that the exponent is not a constant and so cannot be regarded as a power law.

\section{Case: $\lambda = {\lambda_0\Pi}$ and $\sigma=\sigma_0$}

As before, we first would like to write $\lambda$ as purely a function of $S(t)$:
\begin{equation}
    \Pi=\Pi_0 e^{\lambda_0(S -S_0)}.
\end{equation}
This means that $\lambda = \lambda_0 \Pi_0 e^{\lambda_0(S -S_0)} $, and $\lambda' = \lambda_0 \Pi_0$. For this case, we need to rederive the differential equation. For this case, we have:
\begin{equation}
    \begin{cases}
        \int_0^tdt  \lambda\frac{dS}{dt} S(t) = \Pi_0 \left[ \frac{e^{\lambda_0(S-S_0)}(\lambda_0 S - 1)}{\lambda_0} - \frac{\lambda_0 S_0 -1}{\lambda_0} \right] ;\\
        \int_0^tdt  \lambda\frac{dS}{dt}  = \Pi_0 \left[ e^{\lambda_0(S-S_0)} - 1 \right].
    \end{cases}
\end{equation}

One thus obtains another differential equation:
\begin{align}
    \dot{S}(t) &= \frac{b \left[M_0 - \int_0^tdt  \lambda\frac{dS}{dt} S(t) \right]}{\lambda S(t)}  - \frac{p}{\lambda} \left( \Pi_0 +  \int_0^t dt \lambda\frac{dS}{dt} \right)\\
    &= \frac{b \left[M_0 - \Pi_0 \left[ \frac{e^{\lambda_0(S-S_0)}(\lambda_0 S - 1)}{\lambda_0} - \frac{\lambda_0 S_0 -1}{\lambda_0} \right] \right]}{\lambda_0 \Pi_0 e^{\lambda_0(S -S_0)}  S(t)}  - \frac{p}{\lambda_0}\\
    & = - \frac{d}{dS} U(S)
 \end{align}
None of the terms is linear in $S$ and so does not contribute to the power-law exponent of the stationary distribution.

%\clearpage

\section{Application: Numerical Modeling of the Financial Market}\label{sec: application}
In this section, we apply the proposed model to simulate an artificial stock market. The general set of difference equations we rely on is
\begin{equation}
    \begin{cases}
        \Delta M_i(t) = p_i(t)\Pi_i(t) S(t) - b_i(t) M_i(t) \\
        \Delta \Pi_i(t) = - p_i(t) \Pi_i(t) + \frac{b_i(t)M_i(t)}{S(t)}\\
        \Delta S(t) =  \frac{1}{\lambda} \sum_{i=1}^N \Delta \Pi_i(t) + \sigma \epsilon(t).
    \end{cases}
\end{equation}
Note that the difference between this system and the mean-field model we studied analytically is that the stochastic term now directly appear in the third line to model the stochasticity in the price formation process; alternatively, it may also appear at the right of $\Delta M_i$ to model the randomness in the decision-making process of the participants. There is no reason to think that this set of equations may simplify to an equation similar to Eq. (13); when the stochastic term appears directly in the set of equations, the model is no longer analytically tractable. More complicated dynamics may arise when the agents are heterogeneous. Moreover, this model's underlying dynamics are discrete-time, and it is unclear whether the continuous-time approximation will still hold throughout the dynamics. In fact, it is now well-established that discrete-time dynamics can be much more complicated than the continuous-time counterpart, often leading to unpredictability and chaos \cite{may1976simple}. With the additional sources of complications, it is reasonable to investigate the simplest kind of the dynamics offered by the simulation, where we set $\lambda=\lambda_0$ and $\sigma=\sigma_0$ to be constant.

\begin{figure}[t!]
    \centering
    \includegraphics[width=0.4\linewidth]{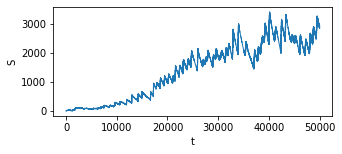}
    \includegraphics[width=0.4\linewidth]{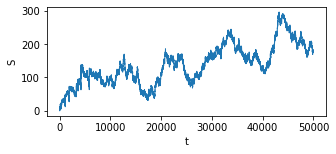}
    
    \includegraphics[width=0.4\linewidth]{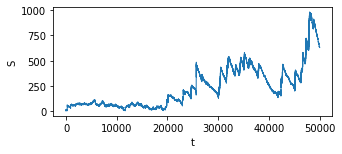}
    \includegraphics[width=0.4\linewidth]{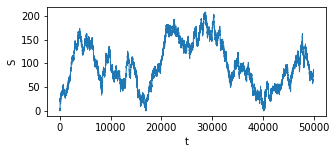}
    \caption{Examples of the simulated price trajectories with different seeds. Even in this simplest case, the generated trajectories are complex and rather realistic.}
    \label{fig:example trajectories}
\end{figure}

\begin{figure*}[t!]
    \centering
    %\begin{subfigure}{0.48\linewidth}
    %\includegraphics[\linewidth]{plots/new_experiments/wealth-distribution.png}
    %\caption{W}
    %\end{subfigure}
    %\begin{subfigure}{0.3\linewidth}
    \includegraphics[width=0.32\linewidth]{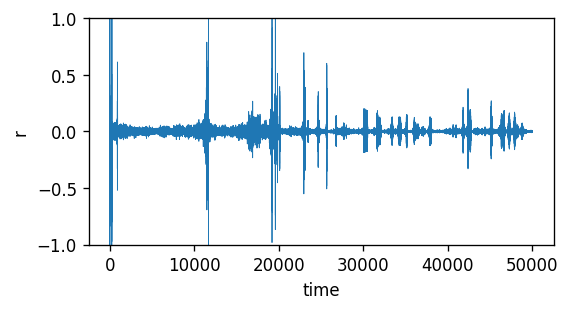}
    %\caption{volatility clustering}
    %\end{subfigure}
    %\begin{subfigure}{0.3\linewidth}
    \includegraphics[width=0.25\linewidth]{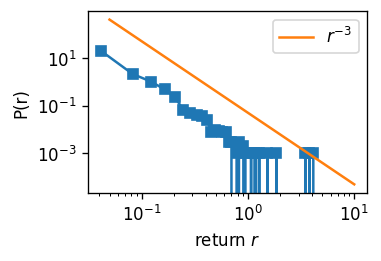}
    %\caption{Heavy tail of return}
    %\end{subfigure}
    %\begin{subfigure}{0.3\linewidth}
    \includegraphics[width=0.25\linewidth]{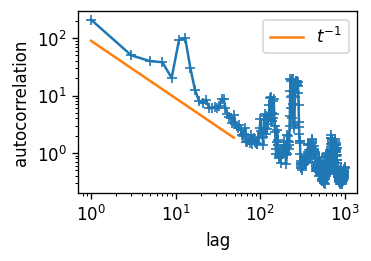}
    %\caption{Power law decay of $c(t)$}
    %\end{subfigure}
    \caption{Measured stylized facts in a simulated market with heterogeneous agents. Left: the volatility clustering effect; Middle: the heavy-tail distribution of return; Right: the power-law decay of the correlation of the return in time, $c(t):= \mathbb{E}_k\mathbb{E}[r_{k+t}^2 r_k^2 ]$ (lower right); The measured values agree with the established empirical values. See a summary in \cite{cont2001empirical}.}
    \label{fig: stylized facts}
\end{figure*}

To make the models heterogeneous in the simplest way, we assign fixed but different $p,\ b$ to the agents, drawn from a bounded distribution with a mean $0.5$. We set $\sigma=1$ and $\lambda=N=1000$. To model the effect of taxation, we decay the cash of every agent by $0.1\%$ at every time step \footnote{Numerically, we observe that this trick also stabilizes the dynamics and prevents divergence.}.
See Figure~\ref{fig:example trajectories} for examples of the trajectory with different seeds. We see that the model alternates between sharp peaks and relatively stable low-price periods, mimicking the business cycles observed in real markets. In the theory section, we discussed stylized facts related to the trading volume and wealth. Here, we experimentally study whether this model exhibits other stylized facts such as volatility clustering and power laws in return. A summary of the standard related stylized facts can be found in \cite{cont2001empirical}; the statistics that are of financial significance are related to the return $r_t :=\log(S_t) - \log(S_{t-1})$. In particular, many statistical facts about $r_t$ are universal, appearing across different markets, time and countries. For example, $r$ is shown to be a heavy-tail distribution with clearly defined power-law index; when $|r_t|$ is large, $|r_{t+1}|$ is also likely to be large, and this effect is often called ``volatility clustering"; quantitatively, one can measure the autocorrelation $c_t:=\mathbb{E}_k\mathbb{E}[r_{k+t}^2 r_k^2 ]$ of $r_t$ and, it is shown that, the decay of $c_t$ in time also follows a power-law form. See Figure~\ref{fig: stylized facts}. We see that even in this simple model of interacting heterogeneous agent simulation, various realistic forms of stylized facts are reproduced. This shows that the proposed framework is flexible enough to be extended to model real problems, which we plan to investigate in the future.

%\bibliographystyle{plain}
%\bibliography{ref}